\documentclass[aps,prl,twocolumn,superscriptaddress,showpacs,preprintnumbers,amsmath,amssymb]{revtex4}

\usepackage{graphicx} 
\usepackage{dcolumn}  

\begin{document}

\vspace*{-3\baselineskip}
\resizebox{!}{3cm}{\includegraphics{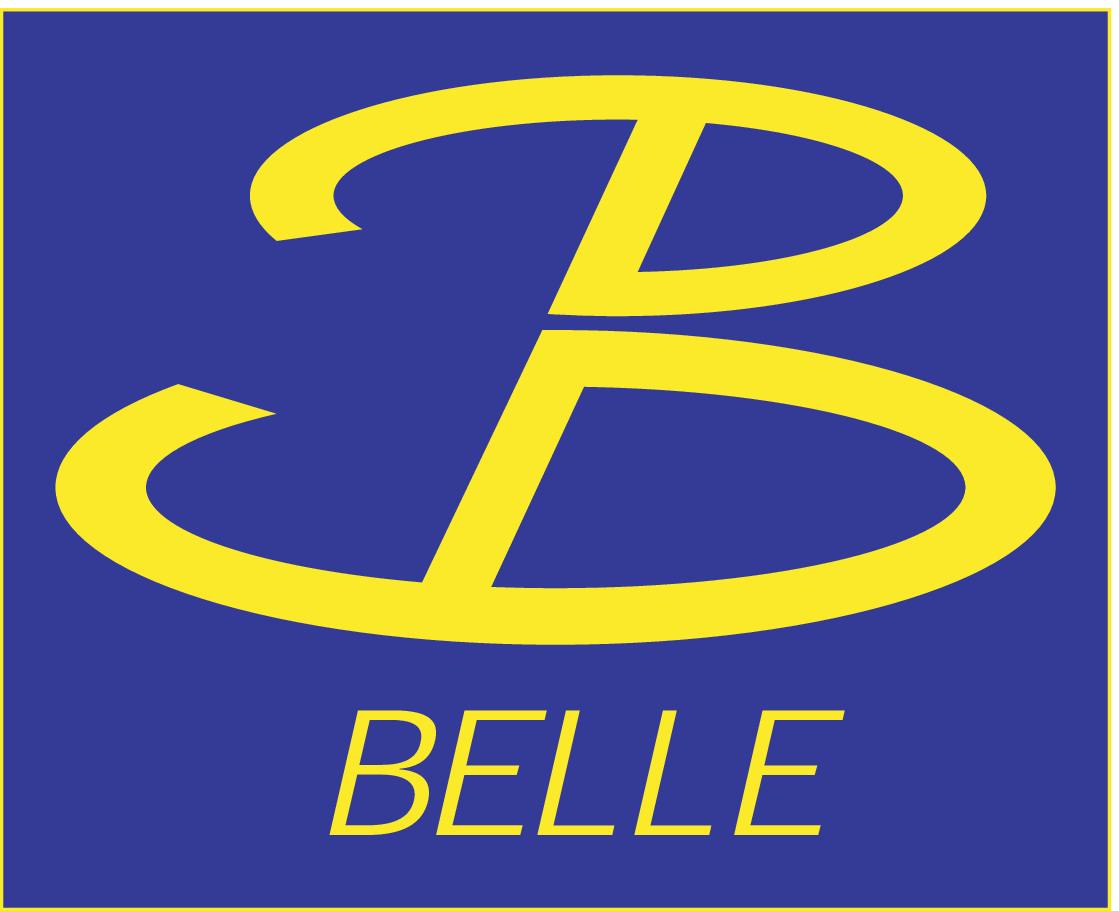}}


\preprint{\vbox{ 
\hbox{   }
\hbox{  }
                  \hbox{\large Belle Preprint 2005-35}
                 \hbox{\large KEK Preprint 2005-87 }
                 \hbox{December 2005/rev. January 2006}
}}

\vspace*{-1.2cm}
\title{ \quad\\[0.5cm]  \ \\ \ \\
Observation of a $\chi'_{c2}$ candidate in 
$\gamma \gamma \to D\bar{D}$ production at Belle }

\begin{abstract}
We report on a search for new resonant states
in the process $\gamma \gamma \to D \bar{D}$.
A candidate
$C$-even charmonium state is observed
in the vicinity of 3.93~GeV/$c^2$.  The production
rate and the angular distribution in the $\gamma\gamma$
center-of-mass frame suggest that this state is the
previously unobserved $\chi'_{c2}$, the $2^3P_2$
charmonium state.
\end{abstract}


\affiliation{Budker Institute of Nuclear Physics, Novosibirsk}
\affiliation{Chiba University, Chiba}
\affiliation{Chonnam National University, Kwangju}
\affiliation{University of Cincinnati, Cincinnati, Ohio 45221}
\affiliation{University of Frankfurt, Frankfurt}
\affiliation{Gyeongsang National University, Chinju}
\affiliation{University of Hawaii, Honolulu, Hawaii 96822}
\affiliation{High Energy Accelerator Research Organization (KEK), Tsukuba}
\affiliation{Hiroshima Institute of Technology, Hiroshima}
\affiliation{Institute of High Energy Physics, Chinese Academy of Sciences, Beijing}
\affiliation{Institute of High Energy Physics, Protvino}
\affiliation{Institute of High Energy Physics, Vienna}
\affiliation{Institute for Theoretical and Experimental Physics, Moscow}
\affiliation{J. Stefan Institute, Ljubljana}
\affiliation{Kanagawa University, Yokohama}
\affiliation{Korea University, Seoul}
\affiliation{Kyungpook National University, Taegu}
\affiliation{Swiss Federal Institute of Technology of Lausanne, EPFL, Lausanne}
\affiliation{University of Maribor, Maribor}
\affiliation{University of Melbourne, Victoria}
\affiliation{Nagoya University, Nagoya}
\affiliation{Nara Women's University, Nara}
\affiliation{National Central University, Chung-li}
\affiliation{Department of Physics, National Taiwan University, Taipei}
\affiliation{H. Niewodniczanski Institute of Nuclear Physics, Krakow}
\affiliation{Niigata University, Niigata}
\affiliation{Nova Gorica Polytechnic, Nova Gorica}
\affiliation{Osaka City University, Osaka}
\affiliation{Panjab University, Chandigarh}
\affiliation{Peking University, Beijing}
\affiliation{Princeton University, Princeton, New Jersey 08544}
\affiliation{University of Science and Technology of China, Hefei}
\affiliation{Seoul National University, Seoul}
\affiliation{Shinshu University, Nagano}
\affiliation{Sungkyunkwan University, Suwon}
\affiliation{University of Sydney, Sydney NSW}
\affiliation{Tata Institute of Fundamental Research, Bombay}
\affiliation{Toho University, Funabashi}
\affiliation{Tohoku Gakuin University, Tagajo}
\affiliation{Tohoku University, Sendai}
\affiliation{Department of Physics, University of Tokyo, Tokyo}
\affiliation{Tokyo Institute of Technology, Tokyo}
\affiliation{Tokyo Metropolitan University, Tokyo}
\affiliation{Tokyo University of Agriculture and Technology, Tokyo}
\affiliation{University of Tsukuba, Tsukuba}
\affiliation{Virginia Polytechnic Institute and State University, Blacksburg, Virginia 24061}
\affiliation{Yonsei University, Seoul}
   \author{S.~Uehara}\affiliation{High Energy Accelerator Research Organization (KEK), Tsukuba} 
   \author{K.~Abe}\affiliation{High Energy Accelerator Research Organization (KEK), Tsukuba} 
   \author{K.~Abe}\affiliation{Tohoku Gakuin University, Tagajo} 
   \author{I.~Adachi}\affiliation{High Energy Accelerator Research Organization (KEK), Tsukuba} 
   \author{H.~Aihara}\affiliation{Department of Physics, University of Tokyo, Tokyo} 
  \author{K.~Arinstein}\affiliation{Budker Institute of Nuclear Physics, Novosibirsk} 
   \author{Y.~Asano}\affiliation{University of Tsukuba, Tsukuba} 
   \author{V.~Aulchenko}\affiliation{Budker Institute of Nuclear Physics, Novosibirsk} 
 \author{T.~Aushev}\affiliation{Institute for Theoretical and Experimental Physics, Moscow} 
   \author{A.~M.~Bakich}\affiliation{University of Sydney, Sydney NSW} 
% ...
\collaboration{The Belle Collaboration}
\noaffiliation

\maketitle

\tighten

{\renewcommand{\thefootnote}{\fnsymbol{footnote}}}
\setcounter{footnote}{0}

  The masses and other properties of 
the ground and excited states
of charmonium provide valuable input to QCD models
that describe heavy quarkonium systems.  To date, 
radial excitation states of charmonium 
are established only for
the $^{2S+1}L_J =$ $^3S_1$ ($\psi$) and, recently, the
$^1S_0$ ($\eta_c$)~\cite{etac2s} states. Although the lowest
$^3P_J$ states ($\chi_{cJ}$) are already well
established, their radial excitations have not yet
been observed.

The first radially excited $\chi_{cJ}$ states
are predicted to have masses between
3.9 and 4.0~GeV/$c^2$~\cite{godfrey,eichten}, which is
considerably above $D\bar{D}$ 
threshold.  If the masses of these states lie between
the $D\bar{D}$ and $D^*\bar{D}^*$ thresholds, the 
$\chi_{c0}(2P)\ (\chi'_{c0})$  and $\chi_{c2}(2P)\ (\chi'_{c2})$
are expected to decay primarily into $D\bar{D}$, although 
the $\chi'_{c2}$ could also decay to $D\bar{D}^*$ if it is
energetically allowed.
(The inclusion of charge-conjugate reactions
is implied throughout this paper.)
Recently,  two new charmonium-like states in this mass region, 
the $X(3940)$~\cite{belledccbar} and $Y(3940)$~\cite{y3940}, 
were reported by Belle.  Neither of these states 
has been observed to decay to $D\bar{D}$~\cite{belledccbar}.

In this paper we report on a search for the 
$\chi'_{cJ}$ ($J=0$ or 2) states
and other $C$-even charmonium states
in the mass range of 3.73 - 4.3~GeV/$c^2$
produced via the process $\gamma \gamma \to D\bar{D}$.

The analysis uses data recorded in
the Belle detector at the KEKB $e^+e^-$ asymmetric-energy 
(3.5 on 8 GeV) collider~\cite{kekb}.
The data sample corresponds to 
an integrated luminosity of 395~fb$^{-1}$,
accumulated on the 
$\Upsilon(4S)$ resonance $(\sqrt{s} = 10.58~{\rm GeV})$
and 60~MeV below the resonance.
We study the two-photon process
$e^+e^-\rightarrow e^+e^- D\bar{D}$
in the ``zero-tag'' mode,  where  
neither the final-state electron nor positron is detected,
and the $D\bar{D}$ system has very small transverse momentum. 

A comprehensive description of the Belle detector is
given elsewhere~\cite{belle}.
Charged tracks are reconstructed in a central
drift chamber (CDC) located in a uniform 1.5~T solenoidal magnetic field.
The $z$ axis of the detector and the solenoid are along the positron beam,
with the positrons moving in the $-z$ direction.  
Track trajectory coordinates near the
collision point are measured by a
silicon vertex detector (SVD).  Photon detection and
energy measurements are provided by a CsI(Tl) electromagnetic
calorimeter (ECL).  
Silica-aerogel Cherenkov counters (ACC) provide separation between kaons 
and pions for momenta above
1.2~GeV/$c$.  The time-of-flight counter (TOF) system consists 
of a barrel of 128 plastic scintillation counters, and is 
effective for $K/\pi$ 
separation for tracks with momenta below 1.2~GeV/$c$. 
Low energy kaons are also identified by specific
ionization ($dE/dx$) measurements in the CDC.

Kaon candidates are separated from pions based on normalized kaon and pion 
likelihood functions obtained from the particle identification system 
($L_K$ and $L_{\pi}$, respectively) 
with a criterion, $L_K/(L_K+L_{\pi})>0.8$, which gives a typical 
identification efficiency of 90\% with a probability of 3\%
for a pion to be misidentified as a kaon.
All tracks that are not identified as kaons are treated as
pions. 

Signal candidates are triggered by a variety
of track-triggers that require two or more CDC 
tracks with associated TOF hits, ECL clusters or a minimum
sum of energy in the ECL.  For the four and six charged
track topologies used in this analysis,
the trigger conditions are complementary to each other
and, in combination, provide a
high trigger efficiency, ($96\pm3$)\%.

We search for exclusive $D\bar{D}$ production in
the following four combinations of decays:
\begin{eqnarray}
\gamma\gamma &\to& D^0\bar{D}^0,\ D^0 \to K^-\pi^+,\ \bar{D}^0 \to K^+\pi^-\ \
\ \ \ \ \ \ \ \ {\rm (N4)}, \nonumber \\
\gamma\gamma &\to& D^0\bar{D}^0,\ D^0 \to K^-\pi^+,\ \bar{D}^0 \to K^+\pi^-\pi^0
\ \ \ \ \ \ \ {\rm (N5)}, \nonumber \\
\gamma\gamma &\to& D^0\bar{D}^0,\ D^0 \to K^-\pi^+,\ \bar{D}^0 \to K^+\pi^-\pi^+
\pi^-\ \ \ {\rm (N6)}, \nonumber \\
\gamma\gamma &\to& D^+D^-,\ D^+ \to K^-\pi^+\pi^+,\ D^- \to K^+\pi^-\pi^- {\rm (C6)}. \nonumber
\end{eqnarray}
The symbols in parentheses are used to designate
each of the final states.
For the four-prong processes (N4 and N5) the selection criteria are:
four charged tracks, each one with (L) a transverse momentum 
to the $z$ axis in the laboratory frame 
of $p_t >0.1$~GeV/$c$;
two or more tracks must have (S) $p_t >0.4$~GeV/$c$
and $17^\circ < \theta <150^\circ$, where $\theta$ is
the laboratory frame polar angle; 
no photon clusters with an energy greater than 400~MeV; 
the charged track system consists of a 
$K^+K^-\pi^+\pi^-$ combination;
the $K^{\pm}\pi^{\mp}$ combination with the larger invariant mass 
should lie within $\pm 15$~MeV/$c^2$ of the nominal $D^0$ mass.
For the N4 process, we require that the 
$K^{\pm}\pi^{\mp}$ combination with the smaller invariant mass be within
$^{+15}_{-20}$~MeV/$c^2$ of the nominal $D^0$ mass.
For the N5 process, we require that the remaining
$K\pi$ combination, when combined with a $\pi^0$ candidate,
has an invariant mass in the range
$1.83$~GeV/$c^2 < M(K^+\pi^-\pi^0) <1.89$~GeV/$c^2$. Candidate 
$\pi^0$'s are formed from pairs of photons with energies greater
than 20~MeV, which fit to 
the $\pi^0\to\gamma\gamma$ hypothesis with $\chi^2 <4$.
If there are multiple $\pi^0$ candidates, we select the one 
that results in $M(K^+\pi^-\pi^0)$ closest to the nominal $D^0$ mass.

For the six-prong processes (N6 and C6), we require exactly
six tracks with particle assignments 
$K^+K^-\pi^+\pi^-\pi^+\pi^-$, where all six pass the looser
track criteria, indicated by (L) above. In addition, either
two to four tracks must pass the more stringent 
track criteria (S) or at least one track has $p_t > 0.5$~GeV/$c$
and the sum of ECL cluster energies is less than 
$0.18\sqrt{s}$, where the cluster energies are measured in
the $e^+e^-$ center-of-mass (c.m.) system.
For the N6 process, one  combination is required to have
$|\Delta M|_1=|M(K^+\pi^-) - m_{D^0}| <15$~MeV/$c^2$ while
the remaining tracks have
$|\Delta M|_2=|M(K^-\pi^+\pi^-\pi^+) - m_{D^0}| <30$~MeV/$c^2$.
When there are multiple combinations, we choose the one with 
the smallest $|\Delta M|_1+|\Delta M|_2$.
For the C6 process, we require 
$|M(K^{\mp}\pi^{\pm}\pi^{\pm}) - m_{D^+}| <30$~MeV/$c^2$
for each of the charge combinations, where
$m_{D^+}$ is the nominal $D^+$ mass.

  For all processes, we require that there
be no extra $\pi^0$ candidates with transverse
momenta larger than 100~MeV/$c$. 
We also apply the following kinematical requirement to the 
$D\bar{D}$ candidate system:
$P_z(D\bar{D}) > (M(D\bar{D})^2-49~{\rm GeV^2}/c^4)/(14~{\rm GeV}/c^3) 
+ 0.6~{\rm GeV}/c$,
where $P_z(D\bar{D})$ and $M(D\bar{D})$ are the momentum component in the
$z$ direction in the laboratory frame and the invariant
mass, respectively.
This condition removes events from initial-state radiation (ISR) 
processes, such as 
$e^+e^- \to D^{(*)}\bar{D}^{(*)} \gamma$, in which
the photon is emitted in the forward direction with respect to
the incident electron.
We compute $M(D\bar{D})$ using the measured 3-momenta of each $D$
candidate ($P_D$) and energy determined
from $E_D = \sqrt{P_D^2 + m_D^2}$, where
$m_D$ is the nominal mass of the neutral or charged
$D$.

The invariant mass distributions for $D$ meson
candidates reconstructed with the above requirements are shown 
in Fig.~1.

We calculate $P_t(D\bar{D})$, the
total transverse momentum in the $e^+e^-$ c.m. 
frame with respect to the 
incident $e^+e^-$
axis that approximates the direction of the
two-photon collision axis. We apply 
the requirement $P_t(D\bar{D}) < 0.05$~GeV/$c$ 
to enhance 
exclusive two-photon $\gamma \gamma \to D\bar{D}$ production.
In the invariant mass region
$M(D\bar{D})<4.3$~GeV/$c^2$,
we find 86 N4-process events,
60 N5-process events, 168 N6-process events
and 128 C6-process events.

\begin{figure}
\centering
\includegraphics[width=8.7cm]{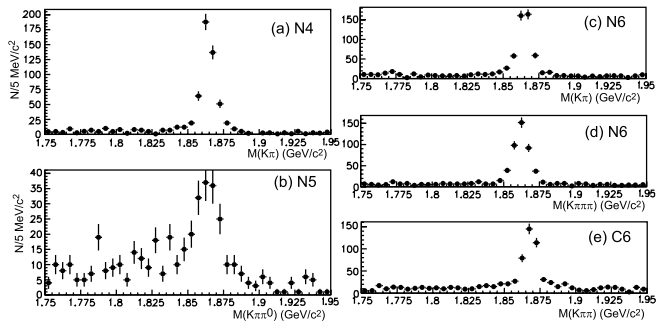}
\label{ddfig1}
\centering
\caption{ Invariant mass distributions of
{\bf (a)} $K^{\mp}\pi^{\pm}$
in N4 candidate events, {\bf (b)} $K^{\pm}\pi^{\mp}\pi^0$ 
in N5 candidate events, {\bf (c)} $K^{\mp}\pi^{\pm}$
in N6 candidate events,  {\bf (d)} $K^{\pm}\pi^{\mp}\pi^{\pm}\pi^{\mp}$ 
in N6 candidate events, and {\bf (e)} $K^{\pm}\pi^{\mp}\pi^{\mp}$
in C6 candidate events. An accompanying $D$ meson candidate 
is required in each event sample.
}
\end{figure}

In Figs.~2(a) and 2(b) we show the  $M(D\bar{D})$ distributions
separately  for $D^0\bar{D}^0$ (sum of N4, N5 and N6) 
and $D^+D^-$.  
The invariant-mass distribution for the 
combined $D^0\bar{D}^0$ and $D^+D^-$ channels is shown
in Fig.~2(c). There, two event concentrations are evident:
one near 3.80~GeV/$c^2$ rather close to the threshold
of $D\bar{D}$ and another near 3.93~GeV/$c^2$.
Each distribution of the four decay combination modes shows
an enhancement near the latter invariant mass.
We apply an unbinned maximum likelihood fit to the combined data
in the region $3.80$~GeV/$c^2 < M(D\bar{D})<4.20$~GeV/$c^2$ using
a relativistic Breit-Wigner signal function 
for the resonant peak near 3.93~GeV/$c^2$ plus a background 
of the form $M(D\bar{D})^{-\alpha}$,
where $\alpha$ is a free parameter.
The invariant mass dependence of the efficiency 
(decreasing by 10\% for an increase of the invariant mass
from 3.80 to 4.20~GeV/$c^2$) and the
two-photon luminosity function are
taken into account in the resonance function.
These are computed using
the TREPS Monte-Carlo(MC) program~\cite{treps}
for $e^+e^- \to e^+e^-D\bar{D}$ production
together with JETSET7.3 decay routines~\cite{lund73} 
for the $D$ meson decays (using 
PDG2004~\cite{pdg} values for the decay branching
fractions). We find from the MC study that the product of the 
efficiency and branching fractions of the two $D$ decay
modes in the $D^+D^-$ channel is about 50\% of that in
the $D^0\bar{D}^0$ channel.

The results of the fit for the resonance mass, width and
total yield of the resonance are $M=3929 \pm 5 (stat)$~MeV/$c^2$,
$\Gamma = 29 \pm 10 (stat)$~MeV and $64 \pm 18 (stat)$~events, respectively.
The mass resolution, which is estimated by MC to be
3~MeV/$c^2$ is taken into account in the fit.
The statistical significance of the peak is $5.3\sigma$, which
is derived from $\sqrt{2\ln(L_{\rm max}/L_0)}$, where $L_{\rm max}$
and $L_0$ are the logarithmic-likelihoods for fits with and 
without a resonance peak component, shown in Fig.~2(c) as solid 
and dashed curves, respectively.

Systematic errors for the parameters $M$ and $\Gamma$ are
2~MeV/$c^2$ and 2~MeV, respectively. The former
is partially due to the uncertainties on the $D$ meson
masses (1~MeV/$c^2$ for the resonance mass). 
We also consider the effect of choosing different Breit-Wigner
functional forms for spin 0 and 2 resonances
and wave functions in this error.


\begin{figure}
\centering
\includegraphics[width=6.5cm]{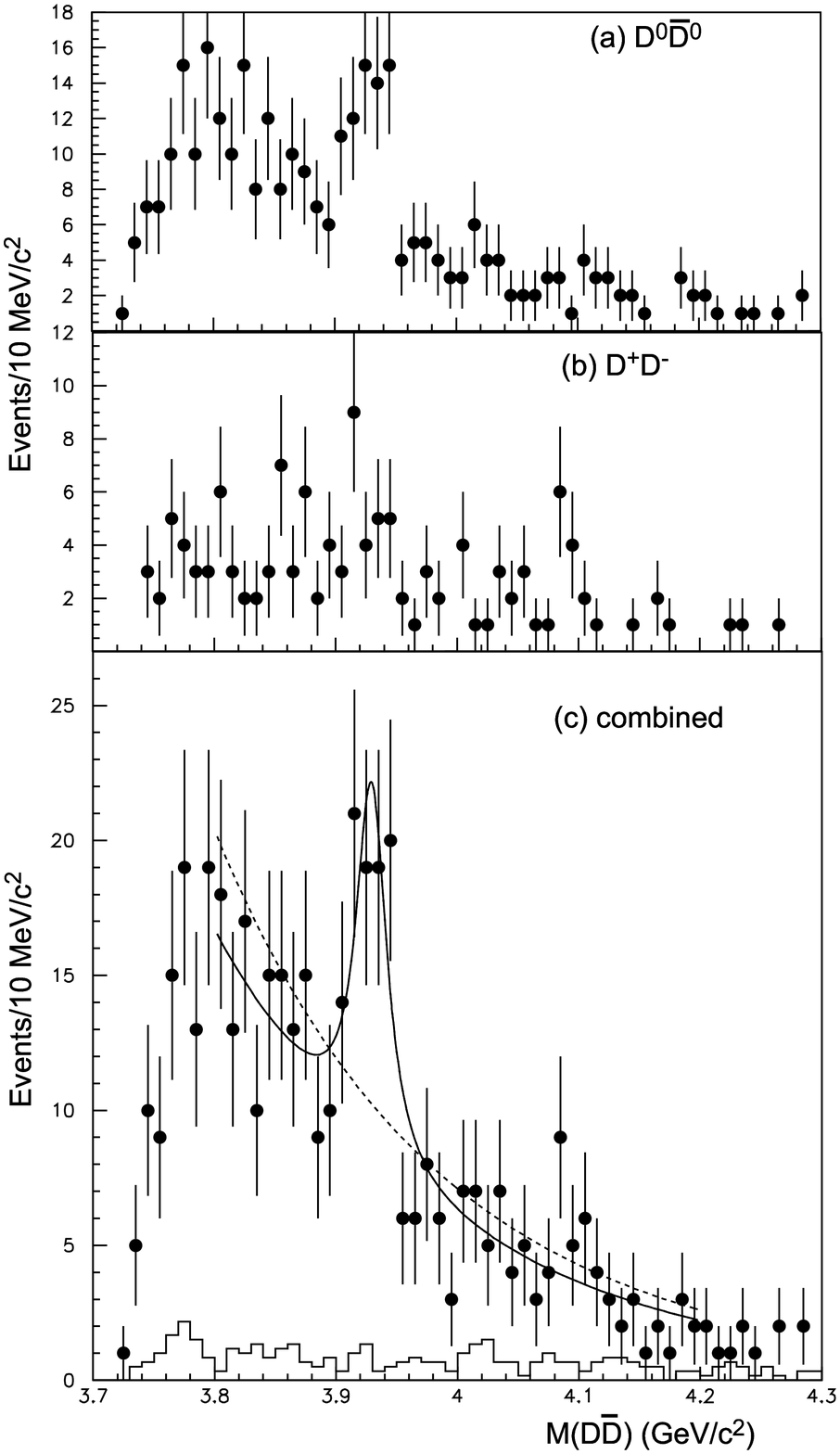}
\label{fig:ddfig5}
\centering
\caption{Invariant mass distributions 
for the {\bf (a)} $D^0\bar{D}^0$ channels 
and {\bf (b)} the $D^+D^-$ mode.  {\bf (c)} The
combined $M(D\bar{D})$ distribution.
The curves show the fits with (solid) and without 
(dashed) a resonance component.
The histogram shows the distribution of the events
from the $D$-mass sidebands (see the text). }
\end{figure}

The $P_t(D\bar{D})$ distribution
in the peak region, $3.91$~GeV/$c^2 < M(D\bar{D}) < 3.95$~GeV/$c^2$,
is shown in Fig.~3.  Here the $P_t$ requirement
has been relaxed.  The experimental
data are fitted by a shape that is
expected for exclusive two-photon $D\bar{D}$ production
plus a linear background.  We expect
non-charm and non-exclusive backgrounds to be nearly linear 
in $P_t(D\bar{D})$.  The fit uses a
binned-maximum likelihood method with the normalizations of the two
components treated as free parameters. The 
linear-background component, 
$1.8 \pm 0.6$ events for $P_t(D\bar{D})
< 0.05$~GeV/$c^2$, and the goodness of 
fit, $\chi^2/d.o.f=14.2/18$, indicate that the events
in the peak region originate primarily from exclusive two-photon
events.

The $P_t(D\bar{D})$ distribution produced
 by $D\bar{D}^*$ and $D^*\bar{D}^*$ events
 is expected to be distorted by the
 transverse momentum of the undetected slow pion(s),
 which peaks around 0.05~GeV/$c$ (dashed histogram in
Fig.~3).
 Such a distortion is not seen
 in the observed $P_t$ distribution.



\begin{figure}
\centering
\includegraphics[width=7cm]{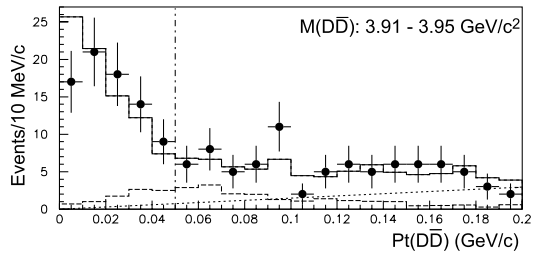}
\label{ddfig6b}
\centering
\caption{Experimental $P_t(D\bar{D})$ distribution 
(points with error bars) for events 
in the $3.91 < M(D\bar{D}) < 3.95$~GeV/$c^2$ region
and the fit (histogram) based on the exclusive 
$\gamma\gamma \to D\bar{D}$
process MC plus a linear background (dotted line). 
The dot-dashed line shows the location of the
$P_t$ selection requirement. The dashed histogram shows 
the expected distribution of the $\gamma \gamma \to D \bar{D}^*$
process followed by $\bar{D}^* \to \bar{D}\pi$ 
with an arbitrary normalization.}
\end{figure}


  We investigate possible backgrounds
from non-$D\bar{D}$ sources using $D$-sideband events.  
The histogram in Fig.~2(c)
shows the invariant mass distribution for events
where the $D$-meson is replaced by a 
hadron system from a 
$D$-signal mass sideband regions above and  below
the signal region with the same width as the signal mass region. 
Here we use two types of sideband events:
one where one $D$-meson candidate is in the signal
mass region, and another where
both entries are from the sidebands.
Since there is no significant event excess in the former type
over the latter, we conclude that
the sideband events are dominated by non-charm backgrounds.
We combine them and appropriately scale in order
to compare to the $D\bar{D}$ signal yield. 
We conclude that 
the candidate events are dominated by $D\bar{D}$ 
(inclusive or exclusive) events in the entire
mass region.

Figures~4(a) and (b) show the $M(D\bar{D})$
distributions for events with
$|\cos \theta^*|<0.5 $  and $|\cos \theta^*|>0.5$,
respectively, where
$\theta^*$ is the angle of a $D$ meson relative 
to the beam axis in the $\gamma\gamma$ c.m. frame.
It is apparent that the events in the 3.93~GeV/$c^2$
peak tend to concentrate at small $|\cos \theta^*|$ values.
The points with error bars in Fig.~4(c) show 
the event yields in the 
$3.91~{\rm GeV/}c^2$ to $3.95~{\rm GeV/}c^2$ 
region versus  $|\cos\theta^* |$.
Background, estimated from events
in the $M(D\bar{D})$ sideband, is
indicated by the histogram. 
The solid curve in Fig.~4(c) shows the expectation
using $\sin^4 \theta^*$ to represent the 
signal from a  spin-2 meson produced with helicity-2 along 
the incident axis~\cite{bellechic2,chic2theo}.
A term proportional to $1 + a\cos^2 \theta^*$ 
that interpolates the  background (dotted  curve) is also 
included.  A small nonuniformity of the signal acceptance 
in the c.m. angle is taken into account.
The comparison to the data has $\chi^2/d.o.f. = 1.9/9.$
Here the functions are normalized
to the numbers of signal and background events
obtained from the fit of the invariant mass distribution,
46 and 33 events, respectively.
A comparison using a constant term to represent the
signal from a spin-0 meson (dashed curve) gives a much
poorer fit: $\chi^2/d.o.f.= 23.4/9$. 
The data significantly favor a spin two
assignment over spin zero.

\begin{figure}
\centering
\includegraphics[width=8.5cm]{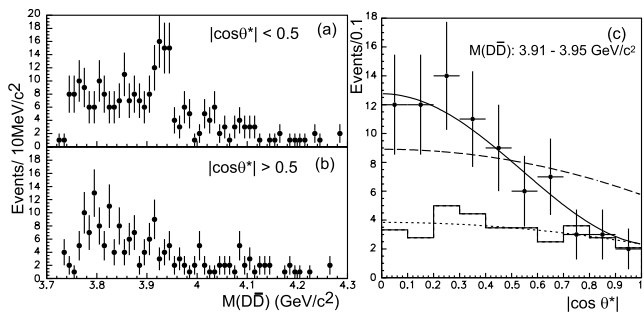}
\label{ddfig7}
\centering
\caption{$M(D\bar{D})$  distributions for
{\bf (a)} $|\cos \theta^*|<0.5 $  and
{\bf (b)} $|\cos \theta^*|>0.5$.
{\bf (c)} The $|\cos \theta^*|$ 
distributions in the $3.91 < M(D\bar{D}) <3.95$~GeV/$c^2$ region
(points with error bars) and background 
scaled from the  $M(D\bar{D})$ sideband (solid histogram). 
The solid and dashed curves are expected distributions  
for the spin two (helicity two) and spin zero hypotheses,
respectively, and contain the non-peak background also
shown separately by the dotted curve.
}
\end{figure}

No charmonium state that decays into $D\bar{D}$ with a
mass near 3.93~GeV/$c^2$ has been previously reported.
This observation  cannot be attributed to a new $J^{PC}=1^{--}$ meson 
($\psi$) produced
by ISR processes, because there are no structures as large as 
the signal at this mass in the  $e^+e^-$ hadronic cross section.

Using the number of observed signal events,
the branching fractions and efficiencies 
for the four decay channels, we determine
the product of the two-photon decay width 
and $D\bar{D}$ branching fraction 
to be
$\Gamma_{\gamma\gamma}(Z(3930)) 
{\cal B}(Z(3930)\to D\bar{D})=0.18 \pm 0.05(stat) \pm 0.03(sys)$~keV,
assuming production of a spin-2 meson.
Here, we define 
$ {\cal B}(Z(3930)\to D\bar{D}) ={\cal B}(Z(3930)\to D^0\bar{D}^0)+
{\cal B}(Z(3930)\to D^+D^-)$ and assume 
${\cal B}(Z(3930)\to D^+D^-) = 0.89
{\cal B}(Z(3930)\to D^0\bar{D}^0)$ according to isospin invariance
and including the effect of the mass difference 
between $D^0$ and $D^\pm$ mesons,
where $Z(3930)$ is used as a tentative designation for
the observed state.

We assign a 17\% total systematic error to the measurement
of the product of the two-photon decay width and the branching fraction,
as shown in the above result.
This is primarily due to uncertainties in the track reconstruction 
efficiency (7\%), selection efficiency (8\%), kaon identification (4\%), 
choice of the fit function
and background shape (5\%), luminosity function (5\%), 
and the $D$-meson branching fractions (9\%), added in quadrature with
other smaller factors.

The observed signals for the $D^{0}\bar{D}^{0}$ and $D^+D^-$ modes 
are consistent with isospin invariance. The ratio of the branching
fractions is measured to be 
${\cal B}(Z(3930) \to D^+D^-)/{\cal B}(Z(3930) \to D^0\bar{D}^0)=
0.74 \pm 0.43 (stat) \pm 0.16(sys)$.
The results on mass, decay angular distributions
and $\Gamma_{\gamma\gamma}{\cal B}(\to D\bar{D})$~
are all  consistent with expectations for the 
$\chi'_{c2}$, the $2^3P_2$ charmonium state~\cite{godfrey, eichten, muenz}.

  In summary, we have observed an enhancement in $D\bar{D}$ invariant mass 
near 3.93~GeV/$c^2$ in $\gamma \gamma \to D\bar{D}$  
events.  The statistical significance of the signal
is $5.3\sigma$.
The observed angular distribution 
is consistent with two-photon production of
a tensor meson.
Results for the mass, width, and the product of
the two-photon decay width times the branching fraction to $D\bar{D}$ 
are: $M=3929 \pm 5(stat) \pm 2(sys)$~MeV/$c^2$, 
$\Gamma = 29 \pm 10(stat) \pm 2(sys)$~MeV
and $\Gamma_{\gamma\gamma}{\cal B}(\to D\bar{D})=0.18 \pm 0.05(stat) \pm 
0.03(sys)$~keV (assuming $J=2$), respectively.  The
measured properties are consistent with expectations
for the previously unseen $\chi'_{c2}$ 
charmonium state.

We thank the KEKB group for excellent operation of the
accelerator, the KEK cryogenics group for efficient solenoid
operations, and the KEK computer group and
the NII for valuable computing and Super-SINET network
support.  We acknowledge support from MEXT and JSPS (Japan);
ARC and DEST (Australia); NSFC (contract No.~10175071,
China); DST (India); the BK21 program of MOEHRD, and the
CHEP SRC and BR (grant No. R01-2005-000-10089-0) programs of
KOSEF (Korea); KBN (contract No.~2P03B 01324, Poland); MIST
(Russia); MHEST (Slovenia);  SNSF (Switzerland); NSC and MOE
(Taiwan); and DOE (USA).

\end{document}